\documentclass[lettersize,journal]{IEEEtran}
\usepackage{amsmath,amsfonts}
\usepackage{mathtools}
\usepackage{algorithmic}
\usepackage{algorithm}
\usepackage{array}
\usepackage[caption=false,font=normalsize,labelfont=sf,textfont=sf]{subfig}
\usepackage{textcomp}
\usepackage{stfloats}
\usepackage{url}
\usepackage{diagbox}
\usepackage{multirow}
\usepackage{makecell}
\usepackage{verbatim}
\usepackage{graphicx}
\usepackage{cite}
\begin{document}

\title{LSR-Net: A Lightweight and Strong Robustness Network for Bearing Fault Diagnosis in Noise Environment}

\author{Junseok Lee,
        Jihye Shin,
        Sangyong Lee,
        *Chang-Jae Chun
        % <-this % stops a space
\thanks{* Corresponding author}}

\markboth{Preprint}
{Shell \MakeLowercase{\textit{J. Lee et al.}}: LSR-Net: A Lightweight and Strong Robustness Network for Bearing Fault Diagnosis in Noise Environment}
% \markboth{IEEE TRANSACTIONS ON INSTRUMENTATION AND MEASUREMENT, VOL. xx, 20xx}%


\maketitle

\begin{abstract}
Rotating bearings play an important role in modern industries, but have a high probability of occurrence of defects because they operate at high speed, high load, and poor operating environments. Therefore, if a delay time occurs when a bearing is diagnosed with a defect, this may cause economic loss and loss of life. Moreover, since the vibration sensor from which the signal is collected is highly affected by the operating environment and surrounding noise, accurate defect diagnosis in a noisy environment is also important. In this paper, we propose a lightweight and strong robustness network (LSR-Net) that is accurate in a noisy environment and enables real-time fault diagnosis. To this end, first, a denoising and feature enhancement module (DFEM) was designed to create a 3-channel 2D matrix by giving several nonlinearity to the feature-map that passed through the denoising module (DM) block composed of convolution-based denoising (CD) blocks. Moreover, adaptive pruning was applied to DM to improve denoising ability when the power of noise is strong. Second, for lightweight model design, a convolution-based efficiency shuffle (CES) block was designed using group convolution (GConv), group pointwise convolution (GPConv) and channel split that can design the model while maintaining low parameters. In addition, the trade-off between the accuracy and model computational complexity that can occur due to the lightweight design of the model was supplemented using attention mechanisms and channel shuffle. In order to verify the defect diagnosis performance of the proposed model, performance verification was conducted in a noisy environment using a vibration signal. As a result, it was confirmed that the proposed model had the best anti-noise ability compared to the benchmark models, and the computational complexity of the model was also the lowest. In addition, as a result of measuring the inference and delay time at the edge device where defect diagnosis is mainly performed, the lowest delay time and the fastest inference time of the proposed model were confirmed.
\end{abstract}


\section{Introduction}
\IEEEPARstart{R}{otating} bearings are widely used components in rotating machines such as aircraft engines and high-speed railways, and have a significant impact on the performance and operating efficiency of the machines \cite{qin_20,song_18}. However, because bearings operate at high speeds, high loads, and harsh working environments, the probability of defects is high \cite{Wagner_20}. As a representation of this, 45 to 55\% of system and machine failures occur due to bearing defects \cite{lanham_02,nandi_05}. Therefore, bearing condition monitoring and defect diagnosis are important to ensure normal operation of rotating machinery systems and to prevent economic losses and human casualties that may occur due to bearing defects \cite{fang_21,cui_22,du_16}.
In order to diagnose bearing defects, it is necessary to analyze characteristics and patterns related to the defect using signals collected from sensors, but it takes a lot of time and money to analyze and diagnose these directly by humans \cite{zhang_20}. Therefore, bearing fault diagnosis research was conducted using classical machine learning (ML) algorithms such as Bayesian classifier, Artificial Neural Network (ANN), Support Vector Machine (SVM), and Principle Component Analysis (PCA) \cite{zhang_20,lu_07,yu_06}. However, this classical ML has the disadvantage of requiring a lot of prior knowledge and experience in signal processing and classical ML algorithms \cite{wang_20}. On the other hand, deep learning (DL) models require less domain-related prior knowledge than classical ML and have better feature extraction capabilities than classical ML \cite{zhu_23}. Therefore, recently, methods that utilize the advantages of DL have been studied \cite{wang_20,jia_22}. In particular, studies using convolutional neural networks (CNN) have achieved surprising success in bearing fault diagnosis tasks \cite{fu_23,fang_23}.

Recently, the importance of accurate real-time monitoring has increased due to the development of Industrial Internet of Things (IIoT) technology, which has secured the ability to collect real-time data by installing a condition diagnosis system \cite{he_17,shi_20}. However, because the standard CNN is designed with an accurate but deep structure, it requires high computational complexity of the model and a large number of parameters \cite{fan_24}. Therefore, as the need for models with low storage and computational complexity required for actual monitoring systems has emerged, making CNN models lightweight has become more important \cite{yao_20}. Additionally, the vibration sensor that collects vibration signals expresses the degree to which the machine shakes during operation as a physical displacement value, so the vibration signals are effective data for diagnosing bearing defects \cite{neupane_20}. However, in the case of rotating machines, it is difficult to collect normal signals due to the harsh operating environment and surrounding noise. Therefore, a model that can accurately diagnose faults even for signals affected by noise is needed \cite{Wagner_20,hoang_19}. To solve this problem, previous studies that designed a CNN-based lightweight model capable of real-time fault diagnosis for bearing faults in a noisy environment are as follows.

In \cite{fang_21}, a lightweight model was designed based on Split CNN, and the anti-noise capability was improved by utilizing a vibration extractor designed based on 1D Conv. In \cite{shi_20}, a lightweight model was designed based on depthwise separable convolution (DSConv), and the anti-noise capability was improved compared to standard CNN by using multi-scale learning. In \cite{ma_19}, a lightweight model was designed based on DSConv, and anti-noise capability was improved by utilizing wavelet packet transform (WPT). In \cite{fan_21}, a lightweight model was designed based on DSConv, and anti-noise capability was improved by utilizing attention mechanism (AM) and multi-scale feature learning. In \cite{wang_23}, a lightweight model was designed based on an inverted residual block and self-attention using DSConv, and the anti-noise capability was improved using AM and multi-scale learning. In \cite{hakim_22}, a lightweight model was designed using a relatively small number of convolutional layers, and the anti-noise capability was improved using fast Fourier Transform (FFT). In \cite{ma_23}, a model was designed based on group convolution (GConv) and dilated convolution, and the optimal model architecture was found using neural architecture search (NAS). In \cite{fang_23a}, a lightweight model was designed using convolution and self-attention, and the anti-noise capability was improved using absolute value FFT (AVFFT). In \cite{tian_23}, a lightweight model was designed using dilated convolution and max pooling, and the anti-noise capability was improved using AM and multi-sensor data fusion. In \cite{yan_24}, a lightweight model was designed using separable multi-scale convolution and broadcast self-attention blocks, and the anti-noise capability was also improved by utilizing self-attention and multi-scale structures. In \cite{sun_23}, a lightweight model was designed by rearranging the Transformer \cite{vaswani_17} structure, and the anti-noise capability was demonstrated by using the Transformer structure.

Recently, real-time bearing fault diagnosis research has been conducted using the structure of the state-of-the-art (SOTA) model designed based on 2D Convolution, which has proven effective in image classification tasks. In \cite{cui_22,yao_20,wu_20,he_21,xue_22}, a lightweight model was designed using the inverted residual block structure of MobileNetV2, V3 \cite{sandler_18,howard_19}, and anti-noise capability was improved using Gramin angular field (GAF), short-time-frequency transform (STFT), and CWT. In \cite{ling_22}, a lightweight model was designed using patch embedding and DSConv of ViT \cite{dosovitskiy_20}, and anti-noise capability was improved using STFT. In \cite{hu_23,cheng_24}, a lightweight model was designed using the structure of EfficientNet V1, V2 \cite{tan_19,tan_21}, and anti-noise capability was improved using CWT and wavelet transform (WT). In \cite{zhong_22}, a lightweight model was designed using the structure of SqueezeNet \cite{iandola_16}, and anti-noise capability was improved using CWT. In \cite{liu_19}, a lightweight model was designed using the structure of ShuffleNetV2 \cite{ma_18}, and anti-noise capability was improved using STFT.

To summarize previous studies, many studies have used signal processing to improve anti-noise capability in situations where there is noise. However, in the case of such signal processing, many parameter modifications are required to suit each data and model. In addition, multi-scale learning was used to improve fault diagnosis performance. This requires repeated convolution operations to extract feature maps of various scales using convolution, and the real-time performance of fault diagnosis may decrease because the number of feature map channels increases. Therefore, in this paper, we design and propose a lightweight and strong robustness network (LSR-Net). To this end, we perform denoising using 1D Convolutional operations that do not require signal processing, and propose a feature enhancement module, denoising and feature enhancement (DFEM), which improves the feature extraction capability by applying activation ensemble to the denoised feature-map. Furthermore, we propose a method to improve the denoising performance by applying pruning to the 1D Convolutional operations of DFEM in situations that are highly affected by noise. In order to extract features from the feature-maps that passed through DFEM, we propose a Conv-based efficient shuffle (CES) block that can design a model while maintaining low computational complexity by utilizing channel split and attention-based group separable Conv (AGSConv). The contribution of this paper is summarized as follows:
\begin{enumerate}
    \item We propose a lightweight strong robustness network (LSR-Net) that is accurate even in vibration sensor environments that are greatly affected by the operating environment and ambient noise, and that enables real-time fault diagnosis based on the low computational complexity of the model.
    \item We propose a DFEM designed using a denoising module (DM) that denoises raw signals affected by noise and a feature enhancement module (FEM) that performs activation ensemble to more effectively extract features from the denoised feature-map.
    \item We propose a CES block designed using channel split and AGSConv to enable effective and real-time feature extraction from the feature-map passed through the DFEM.
    \item In order to verify whether the proposed model is suitable for actual industrial sites, we provide the results of measuring the model inference time and power consumption on an edge device where fault diagnosis is mainly performed.
\end{enumerate}

\section{Theoretical Background}
\subsection{Group Convolution}
Group Conv (GConv) was proposed in AlexNet \cite{krizhevsky_17} to solve the memory cost problem of GPU \cite{zhang_17}. When performing a standard Conv calculation, if the number of input channels is M and the number of output channels is $N$, the calculation cost of Conv is $M \times N$. However, if the Conv operation of M is performed divided by G (group), which denote the number of groups, the computational cost of GConv is reduced as $\frac{M}{G} \times N$ \cite{huang_18}.
Additionally, GConv enhances performance with fewer parameters by isolating and processing sparse features within each group, as it splits the feature map's channels into separate groups. 
In addition, because GConv performs Conv operation by dividing the channels of the feature-map connected by complex channels into groups, it is easy to find sparse features for each group, which has the advantage of increasing performance with fewer parameters \cite{kim_22}. However, since the Conv operation is performed independently for each group, it can increase both training and inference time. Therefore, GConv should be employed appropriately.

\subsection{Depthwise Separable Convolution (DSConv)}
Since DSConv consists of DConv, which performs convolution operations for each channel, and PConv, which is a $1 \times 1$ kernel size convolution that combines them into one, if a model is designed using this, the computational complexity and size of the model can be reduced compared to a model designed using a general convolution layer \cite{sandler_18,howard_19}. Unlike the standard Conv operation, DConv does not perform Conv operation in the channel direction during operation, but only performs Conv operation in the spatial direction \cite{howard_17}. The computational cost of standard Conv and DConv is calculated as follows \cite{sandler_18}:

\begin{gather}
    \label{eq:Conv}
    Q_{Conv} = D_{K}^{2}MND_{F}^{2} \\
    \label{eq:DConv}
    Q_{DConv} = D_{K}^{2}MD_{F}^{2}
\end{gather}

In Eq. (\ref{eq:Conv}), (\ref{eq:DConv}), $D_{F}^{2}$ is the input feature-map $\mathbf{X} \in \mathbb{R}^{C \times H \times W}$, where the height (H) and width (W) denote a square input feature-map, M is the number of input channels, N is the number of output channels, and $D_{K}^{2}$ denote the size of the kernel used in the Conv calculation. The calculation of the standard Conv in Eq. (\ref{eq:Conv}) receives $D_{F}^{2} \cdot M$ as input. The size of the output feature-map is $D_{F}^{2} \cdot N$, which is parameterized by the kernel size $D_K^2$ of the Conv. In contrast, DConv in Eq. \ref{eq:DConv} can construct layers with lower computational cost because it performs convolution operations for each channel as in. However, since DConv performs Conv operation for each channel, it does not combine each channel to generate a new feature. Therefore, PConv with 1x1 kernel size is used to combine each channel to generate a new feature. The computational cost of DSConv, which is designed by combining DConv and PConv, is calculated as follows:

\begin{align}
    Q_{DSConv} = D_{K}^{2} M D_{F}^{2} + M N D_{F}^{2}
\end{align}

The computational complexity ratio of standard Conv and DSConv can be compared as follows:

\begin{align}\label{eq:compare Conv and DSConv}
      \begin{split}
         & \frac{Q_{DSConv} = D_{K}^{2}  M  D_{F}^{2} + M  N  D_{F}^{2} }{ Q_{Conv} = D_{K}^{2}  M  N  D_{F}^{2}} = \frac{1}{N} + \frac{1}{D_{K}^{2}} 
      \end{split}
\end{align} 

For example, if $N$ is 128 and $D_K$ is 3, using DSConv, we can see that the computational cost is reduced by $\cong$ 11.89\% compared to standard Conv. However, as the number of groups increases during Conv operation, a high memory access cost (MAC) is expected, which causes a bottleneck in the GPU or CPU. MAC is calculated as follows:

\begin{align}\label{eq:MAC}
      \begin{split}
        & MAC = D_{K}^{2}  D_{F}^{2}  (M  N) + \frac{M  N}{g}\\
        & = D_{K}^{2}  D_{F}^{2}  M + \frac{B  g}{M} + \frac{B}{D_{F}^{2}} 
      \end{split}
\end{align} 

Here, $B=(D_{F}^{2} M  N)/g$ stands for floating-point operations (FLOPs). When comparing the computational cost by fixing the input value $M D_{F}^{2}$, it can be confirmed that the size of MAC increases as the number of g increases. When adjusting the number of groups based on the existing PConv and comparing the inference speed, it was confirmed that the inference was about 35.6\% faster when $g=1$ than when $g=8$ \cite{ma_18}.

\subsection{Attention Mechanism}
Attention mechanism is a method of giving a higher influence when highly relevant information is presented within a feature-map and a lower influence when less relevant information is presented \cite{guo_22}. 
The learning performance of the model may be improved by efficiently readjusting the feature-map without increasing the computational complexity by utilizing such an attention mechanism. Representative attention mechanisms include squeeze-and-excitation (SE) block \cite{hu_18} and spatial attention module (SAM) \cite{qin_19}. The SE block is a method of applying attention to the area of the channel. When performing the Conv operation, the information on the complicated channels is very important. Therefore, the SE block compresses the area of the feature-map into a channel by applying global average pooling (GAP), and refines important features of the channel by using the fully-connected (FC) layer and the non-linearity (activation function). The process by which the input feature-map $\textbf{X} \in \mathbb{R}^{C \times H \times W}$ is input into the SE block and becomes refined feature-map.

\begin{gather}
    \mathbf{X}^{C} = GAP(\mathbf{X}) \\
    \mathbf{X}^{C/r} = ReLU(FC_{1}(\mathbf{X}^C)) \\
    \hat{\mathbf{X}}^{C} = sigmoid(FC_{2}(\mathbf{X}^{C/r})) \\
    \mathbf{X}' = \textbf{X} \cdot \hat{\mathbf{X}}^{C} 
\end{gather}

SAM is a method of applying attention to the spatial region of the feature-map, as opposed to SE block. Contrary to SE block, SAM applies average pooling (AP) to the input feature-map to compress the feature-map into a spatial domain and refine important spatial features using Conv operation and sigmoid. The process by which input feature-map $\textbf{X} \in \mathbb{R}^{C \times H \times W}$ is input into SAM and becomes refined feature-map

\begin{gather}
    \mathbf{X}^{S} = AP(\mathbf{X}) \\
    \hat{\mathbf{X}}^{S} = sigmoid(Conv(\mathbf{X}^{S})) \\
    \mathbf{X}' = \textbf{X} \cdot \hat{\mathbf{X}}^{S} 
\end{gather}

\section{Proposed Methods}
In this section, we propose a model that enables fast inference time in edge devices and robust defect diagnosis even with high-intensity noise due to the lightweight model design. 
To this end, section \ref{sec:framework} describes the overall defect diagnosis frame work of the proposed model, section \ref{sec:DFEM} describes DFEM, which can denoise and feature enhance signal data containing noise, and section \ref{sec:CESB} describes the detailed structure of CES block and LSR-Net, which extracts feature-maps that have passed through the section \ref{sec:DFEM} method.

\subsection{Proposed Lightweight Network}\label{sec:framework}

The structure of the LSR-Net, the lightweight network proposed in this paper, is shown in Fig. \ref{fig:proposed framework}. In Fig. \ref{fig:proposed framework}, "BN" denote batch-norm, "HS" denote hard swish, and "GAP" denote global average pooling. 

The input data of LSR-Net is a raw signal $\textbf{x}_0 \in \mathbb{R}^{1 \times N}$ with length N, and goes through the denoising and feature enhancement processes using DFEM. After that, the raw signal is converted to a 2D matrix of 3 channels, $\textbf{F} \in \mathbb{R}^{3 \times H \times W}$, where H and W represent the height and width, respectively. In order to align and compress F, which has different information due to various nonlinearities, a 2D conv with a receptive field of 3x3 size and average pooling are utilized. The sorted and compressed feature-map was used to perform feature extraction K times using the CES block. Since the extracted feature-map is in the form of a 2D matrix, it must be converted to a 1D matrix for fault diagnosis. In this process, flattening a general 2D matrix can greatly increase the computational complexity of the FC layer \cite{hsiao_19}. Therefore, the 2D matrix from which features are extracted is compressed into a single vector using global average pooling (GAP) and used as input to the FC layer. Defect diagnosis is performed using $\hat{y} \in \mathbb{R}^{1 \times O}$ that passed through the FC layer. $O$ represents the number of defect types to be predicted.

\subsection{Denoising and Feature Enhancement Module (DFEM)}\label{sec:DFEM}
\begin{figure}[!htb]
    \centering
    \includegraphics[width=0.55\columnwidth]{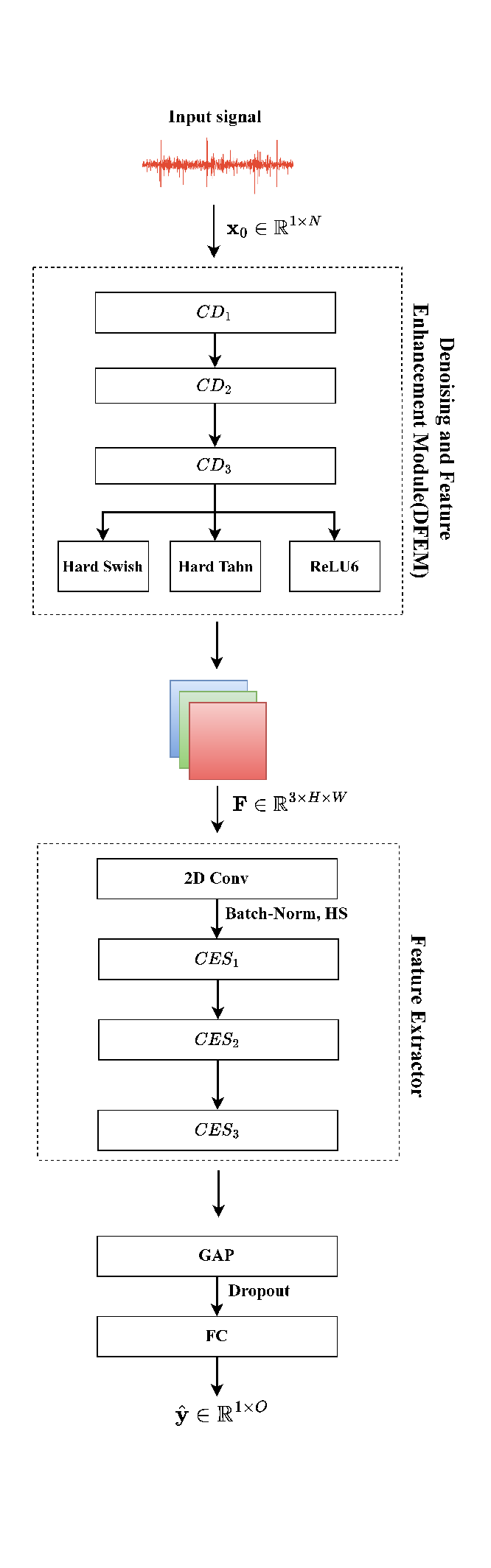}
    \caption{Proposed lightweight network framework}
    \label{fig:proposed framework}
\end{figure}

When vibration signals are collected in actual work sites, they are affected by noise due to external factors. To solve this problem, this paper proposes DFEM. DFEM is configured as shown in Fig. \ref{fig:proposed framework}. In \ref{sec:DM}, DM, which denoises raw signals by utilizing $n$ Conv-based denoising (CD) blocks, is described. In \ref{sec:AP-DM}, AP-DM, which can improve anti-noise ability compared to standard DM in situations that are greatly affected by noise by applying adaptive-pruning (AP) to DM, is described. In \ref{sec:FEM}, FEM, which can enhance the characteristics of feature maps by applying various nonlinearities to the denoised feature maps by utilizing activation ensemble, is described.

\subsubsection{Denoising Module (DM)}\label{sec:DM}
\begin{figure}[h]
    \centering
    \includegraphics[height=0.65\columnwidth,width=\columnwidth]{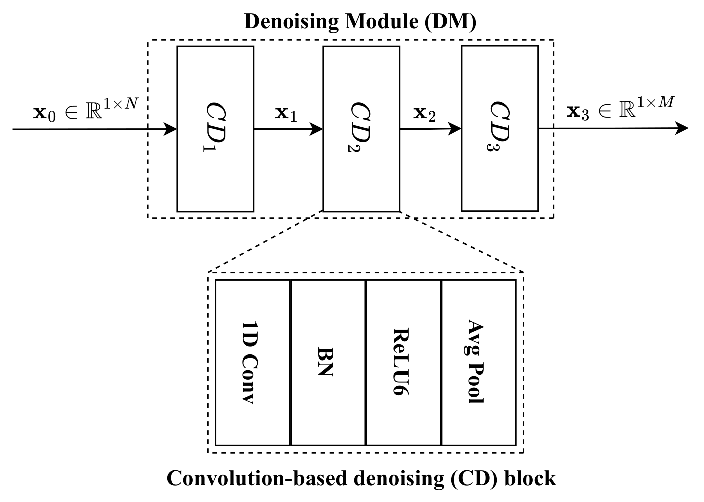}
    \caption{Denoising module (DM).}
    \label{fig:cd_block}
\end{figure}

DM is designed based on CD block as shown in Fig. \ref{fig:cd_block}. The signal $x_0$ affected by noise goes through the denoising process while passing through DM. Feature extraction using 1D Conv of CD block is very effective in denoising raw signal affected by noise because it extracts only important information related to defects among the large amount of information of raw signal \cite{fang_21}. In addition, it is effective in complex pattern recognition by imparting nonlinearity, and it is also effective in terms of lightweight model design because the data dimensional can be effectively reduced by using stride of 1D Convolution and average pooling (AP).

\subsubsection{Adaptive-pruning DM (AP-DM)}\label{sec:AP-DM}
\begin{figure}[h]
    \centering
    \includegraphics[width=0.9\columnwidth]{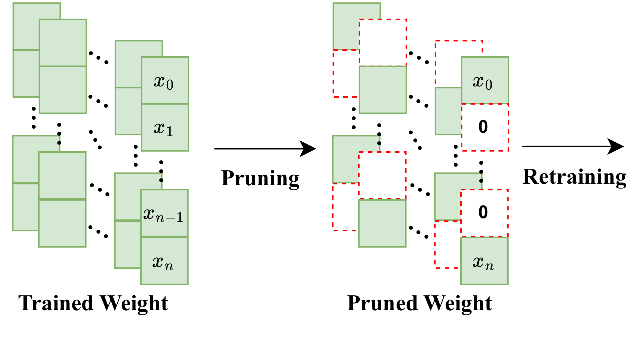}
    \caption{Adaptive pruning-DM (AP-DM).}
    \label{fig:AP-DM}
\end{figure}

When a signal is collected from a sensor, the noise and signal power of the working environment can be measured \cite{fan_22}. Therefore, in this paper, a penalty is applied to the feature extraction process of 1D Conv, which receives raw signals as input in an environment with strong noise power, to improve the denoising ability. The process of AP-DM is as shown in Fig. \ref{fig:AP-DM}. In Fig. \ref{fig:AP-DM}, pruning is applied to the weights of 1D Conv whenever the loss decreases during learning, and the size of some of the learned weights is converted to 0. In this process, the dense feature-map is converted to a sparse feature-map, which can prevent overfitting that can occur in 1D Conv that directly extracts features from data affected by noise. Adaptive Pruning was applied whenever the loss decreased to prevent vanishing gradients that may occur during learning. Also, since the model continuously updates weights, important weights of the model can be restored. When AP was applied to DM, it was confirmed that the denoising ability was improved, enabling more accurate fault diagnosis. Detailed results can be found in Table VI of Section IV.

\subsubsection{Feature enhancement Module (FEM)}\label{sec:FEM}
Fault diagnosis using a single vibration signal has limitations in terms of accuracy, and adopting a signal processing method for this purpose increases the difficulty of the fault diagnosis task because there are many factors to consider, such as the characteristics of the data and parameters for signal processing \cite{huang_22}. Therefore, in this paper, we applied an activation ensemble to a single vibration signal to improve the feature information. HT, ReLU6, and HS were used for the activation ensemble. HT has a normalization effect because it limits the values of neurons in the feature-map to between -1 and 1. In addition, the existing tanh exponent operation requires high computational cost in embedded environments with limited computing power, so it may be unsuitable for bearing fault diagnosis models that are mainly performed in embedded environments \cite{howard_19}. Therefore, HT, which has a relatively low computational cost compared to Tanh, was used. HT is calculated as follows:

\begin{align}
    HT(x) = max(min(x,1),-1)
\end{align}

HT has low computational cost and is effective in normalizing feature maps, but it does not use values below -1 and neuron values above 1 for learning. To solve this problem, ReLU6 was used to supplement it.

HT does not use exponential operation and can be normalized, but it does not use values below -1 and values above 1 for learning. Therefore, it was supplemented by using an activation index that gives various nonlinearity to one feature-map using various activation functions. In this paper, ReLU6 and HS were used for the activation index.

ReLU6 is a function that makes it easy to find a sparse feature with an upper limit of 6 in ReLU ($max(0,x)$) and speeds up the convergence speed \cite{krizhevsky_10}. Therefore, ReLU6 was used to supplement HT with an upper limit of 1 for the value of a neuron used for learning. ReLU6 is calculated as follows:
\begin{align}
    ReLU6(x)= min(max(0, x), 6)
\end{align}

ReLU6 may have a negative effect on learning by causing a dying ReLU phenomenon in which learning is not performed on neurons for negative numbers less than 0. Moreover, the accumulated product of too many negative values can negatively affect learning. Therefore, neurons with bound negative numbers were supplemented by using HS for learning.
HS is a function that replaces the sigmoid function with piece-wise linear hard analog: $ReLU6(x+3) / 6$ in the swish function ($x \cdot sigmoid(x)$) \cite{howard_19}. Through this, faster calculation is possible because the exponential calculation of sigmoid, which requires high calculation cost, is not performed. HS is calculated as follows:

\begin{align}
\label{eq:hard_swish}
    HS(x)=x\cdot\frac{ReLU6(x+3)}{6}
\end{align}

The process by which the input raw signal$ \textbf{x}_0 \in \mathbb{R}^{1 \times N}$ passes through the DFEM to become $\textbf{X}_0 \in \mathbb{R}^{3 \times H \times W}$ is as follows:

\begin{gather}
    \textbf{x}_{n} =  \left\{\begin{matrix}
    DM_n(\textbf{x}_{0}), & SNR \geq 0   \\
    AP-DM_n((\textbf{x}_{0}), & SNR < 0 \\
    \end{matrix}\right.  , n=1,2,3\\
    \textbf{x}' = ED(\textbf{x}_n) \\
    \textbf{F} = \big[ReLU6(\textbf{x}'), HT(\textbf{x}'), HS(\textbf{x}')\big]
\end{gather}

Here, $ED$ denote expand dimension. And since the signal processing was performed using deep learning, additional data preprocessing such as frequency domain analysis is not required \cite{fang_23}. In addition, when the information that gave different nonlinearity to the feature-map is fused, more accurate defect diagnosis performance can be expected by using feature-maps with different information for learning \cite{magar_21}. In this paper, it was confirmed that there is a denoising and feature enhancement effect when the raw signal was converted into a 3-channel 2D matrix using DFEM. Detailed results can be found in Table \ref{tab:DFEM}. 

These ReLU6 and HS were mainly used in the CES block shown in the next section due to their fast convergence speed and low computational cost.

\subsection{Convolution-based Efficient Shuffle (CES) block}\label{sec:CESB}
\begin{figure}[!ht]
    \centering
    \includegraphics[height=0.55\textheight,width=0.9\columnwidth]{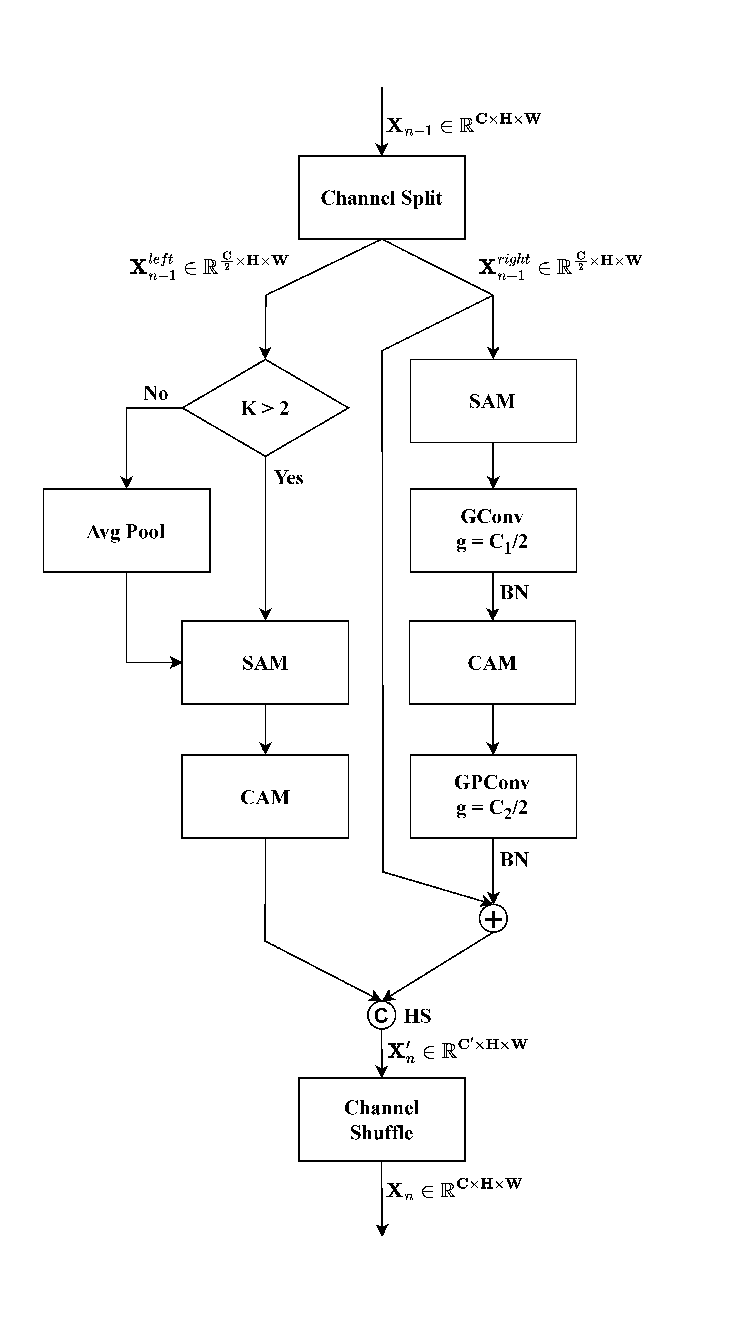}
    \caption{proposed Conv-based efficient shuffle (CES) block}
    \label{fig:CESB}
\end{figure}

Fig. As shown in \ref{fig:proposed framework}, as a result of performing 2D Conv on the output $\textbf{F} \in \mathbb{R}^{3 \times H \times W}$ of DFEM, a 2D matrix $\textbf{X}_0 \in \mathbb{R}^{C \times H \times W}$ of $C$ channels is generated.
Next, in order to feature extract of data of this structure, three CES blocks constructed based on 2D Conv were designed continuously as shown in Fig. \ref{fig:proposed framework}. The specific structure of the $n$th CES block ($n=1,2,3$) is shown in Fig. \ref{fig:CESB}.

In existing studies, DSConv was used to design blocks or layers. However, PConv causes high FLOPs, and DConv is effective in reducing computational complexity, but causes high MAC due to the use of a large number of groups \cite{ma_18}. Therefore, in this paper, GConv and GPConv were used to complementary consider the MAC/FLOPs ratio, which can cause bottlenecks in GPU and CPU, and FLOPs, the computational cost of the model.
Eq. \ref{eq:MAC} for calculate MAC, when $D_k$=3, $D_F$=32, $M=N=64$, the MAC/FLOPs ratio of DConv with number of group 64 is 200\%, while the MAC/FLOPs ratio of GConv with number of group 8 is 200\%. It was confirmed that there was a large difference of 175\% at 25\%. On the other hand, when $D_k=1$, $D_F=32$, $M=64$, $N=128$, the MAC/FLOPs ratio of GPConv with $g=8$ is 18\%, which is lower than the MAC/FLOPs ratio of PConv with $g=1$ (2\%). When compared, it was confirmed that there was a small difference of 16\%. Comparing the FLOPs of Gconv, GPConv and DSConv, it is calculated as follows:

\begin{align}\label{eq:compare proposed and DSConv}
         \frac{Q_{GSConv} = D_{K}^{2} \cdot \frac{M}{g} \cdot M \cdot D_{F}^{2} + \frac{M}{g} \cdot N \cdot D_{F}^{2}}{Q_{DSConv} = D_{K}^{2} \cdot M \cdot D_{F}^{2} + M \cdot N \cdot D_{F}^{2}} 
\end{align} 
For example, if you design a layer or block using Gconv and GPConv when $D_k = 3$, $D_F = 32$, $M = 64$, $N = 128$, $g = 8$, you can design a model with a $\cong 64.2\%$ reduced FLOPs and a small MAC/FLOPs ratio compared to DSConv. In this paper, $C/2$ was used for the number of GConv and GPConv groups.

\begin{table*}[hbt]
    \caption{Structure of the LSR-Net}
    \label{tab:LSR-Net}
    \centering
    {\small
    \resizebox{1.5\columnwidth}{!}{%
        \begin{tabular}{c | c c c c c c}
        \hline\hline 
        Stage & Layer/block    & Kernel size   & Stride   & In Channels & Out Channels & Activation Function   \\ [0.5ex]
        \hline 
        \multirow{5}{*}{DFEM}  & $CD_1$   & $3 \times 1$ & $2 \times 1$ & $1$   & $8$     & ReLU6   \\ [0.5ex]
                               & $CD_2$   & $3 \times 1$ & $2 \times 1$ & $8$   & $16$    & ReLU6   \\ [0.5ex]
                               & $CD_3$   & $3 \times 1$ & $4 \times 1$ & $16$  & $32$    & ReLU6   \\ [0.5ex]
                               & Concatenation   & / & /  & $32$  & $3$    & HT, ReLU6, HS      \\ [0.5ex]
        \hline 
        \multirow{7}{*}{\begin{tabular}[c]{@{}l@{}}Feature\\Extractor\end{tabular}}  
        & 2D Conv & $3 \times 3$ & $2 \times 2$  & $3$ & $16$ & ReLU6  \\  [0.5ex]
        & AP      & $2 \times 2$ & $2 \times 2$ & / & $16$ & HS  \\  [0.5ex]
        & $CES_1$     & $3 \times 3$ & $2 \times 2$ & $16$ & $32$ & HS \\ [0.5ex]
        & $CES_2$     & $3 \times 3$ & $2 \times 2$ & $32$ & $64$ & HS \\ [0.5ex]
        & $CES_3$     & $3 \times 3$ & $1 \times 2$ & $64$ & $64$ & HS \\ [0.5ex]
        & GAP     & /   & / & $64$ & $64$  & / \\ [0.5ex]                                                          
        \hline
        Classifier         & FC        & /        & /             & $64$        & $3$       & SoftMax      \\                   
        \hline\hline
    \end{tabular}
    }}%
\end{table*}

\begin{figure}[ht]
    \centering
    \includegraphics[height=0.4\textheight,width=0.9\columnwidth]{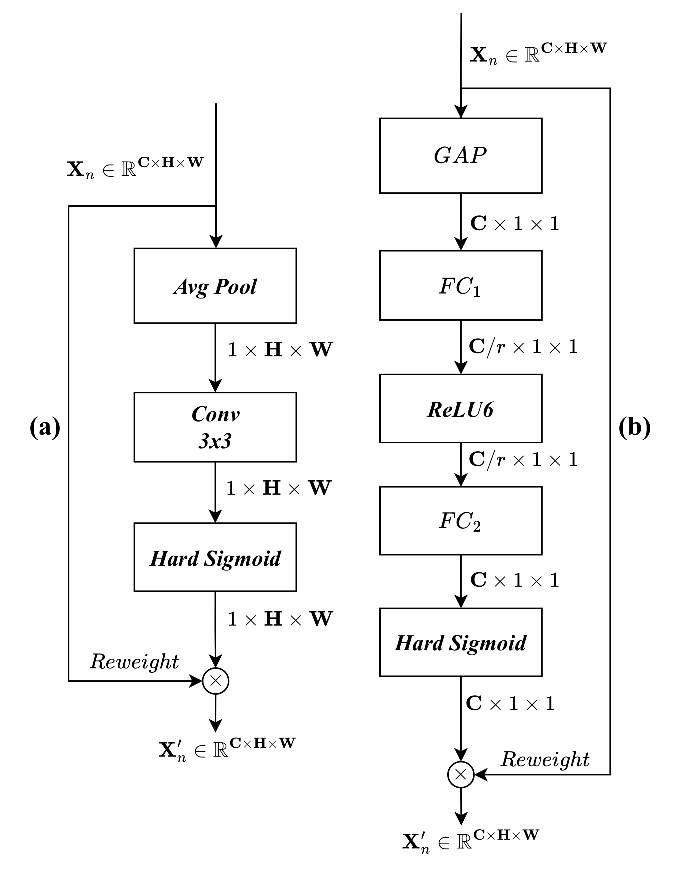}
    \caption{(a) Spatial attention module (SAM) and (b) Channel attention module (CAM).}
    \label{fig:AM}
\end{figure}

In addition, SAM and CAM were applied after GConv and GPConv layers in the CES block, respectively. The structure of SAM and CAM is as shown in Fig. \ref{fig:AM}. 
Since Gconv performed feature extraction for each groups spatial region, SAM was applied, and GPConv performed feature extraction for each groups channel region, CAM was applied to refine the feature-map extracted from each area. Since the CES block is for lightweight design, the sigmoid used for the calculation of the existing attention score in CAM and SAM has been replaced with the hard sigmoid. By not performing the exponential operation of sigmoid, faster inference will be possible at a lower calculation cost in the embedded environment \cite{howard_19}. The hard sigmoid is calculated as follows:

\begin{align}
    f(x) = max(0, min(1, \frac{(x+1)}{2}))
\end{align}

In the CES block, channel splits such as $\textbf{X}_{n-1}^{left} \in \mathbb{R}^{C/2 \times H \times W}$ and $\textbf{X}_{n-1}^{right} \in \mathbb{R}^{C/2 \times H \times W}$ were applied based on channel (C) in input feature-map $\textbf{X}_{n-1} \in \mathbb{R}^{C \times H \times W}$. The channel split maintains the number of channels in the feature-map, but reduces the number of channels used in the Conv operation to reduce computational complexity \cite{ma_18}. In addition, by reusing the previous feature, it is possible to reduce repetitive features used for learning, thereby preventing overfitting, and effectively learning is possible by sharing information learned from the previous layer or block in the next layer \cite{prajapati_21}. In the case of $\textbf{X}_{n-1}^{left}$, feature extraction was performed only on $\textbf{X}_{n-1}^{right}$, leaving it for identity mapping. However, if the stride of GConv is 2 or more, pooling was performed by applying AP to the $\textbf{X}_{n-1}^{left}$ to match the size of the spatial matrix. In addition, it was designed to be concatenated with $\textbf{X}_{n}^{right}$, which performed feature extraction by refining the feature-map by applying SAM and CAM without reusing the previous information as it was. 

GConv has the disadvantage that each output channel performs the Conv operation only on the input channel within the corresponding group \cite{zhang_18}. Therefore, information between different groups cannot be exchanged by continuously performing a Conv operation between the same groups. Therefore, after concatenating $\textbf{X}^{right}$ with final connection \cite{he_16} that may refer to feature extraction and previous information, and $\textbf{X}^{left}$ with AM with concatenation, channel shuffle was applied. It is possible to improve the expressive power of the model when using Gconv by exchanging information between groups by learning different channels instead of the same group \cite{luo_22}. The detailed structure of the LSR-Net designed using DFEM and CES block is shown in Table \ref{tab:LSR-Net}.

The process by which the input feature-map $\textbf{X}_{n-1}$ passes through the CES block to become $\textbf{X}_n$ is as follows:

\vspace{-15pt}
\begin{gather}
    \textbf{X}_{n-1}^{left}, \textbf{X}_{n-1}^{right}=channel\,\,split(\textbf{X}_{n-1}) \\
    \begin{split}
        \textbf{X}_{n}^{right}= & CAM(GPConv(SAM(Conv(\textbf{X}_{n-1}^{right})))) \\
        & + \textbf{X}_{n-1}^{right} 
    \end{split}
\end{gather}

\begin{gather}    
    \textbf{X}_{n}^{left}=  \left\{\begin{matrix}
    CAM(SAM(\textbf{X}_{n-1}^{left})),&s=1  \\
    AP(CAM(SAM(\textbf{X}_{n-1}^{left}))), & s=2 \\
    \end{matrix}\right. \\
    \textbf{X}'_n = Concat\big[\textbf{X}_{n}^{right}, \textbf{X}_{n}^{left}\big] \\
    \textbf{X}_n = Channel\,\,Shuffle(\textbf{X}'_n)
\end{gather}

When designing a block using Channel splits, GConv, and GPConv, it was confirmed that the block could be designed with smaller FLOPs and MACs compared to designing a block using DSConv in the form of a plane. When designing the LSR-Net using the CES block designed in this way, it was confirmed that the inference time was faster than that of the existing benchmark model, and the defect diagnosis accuracy was also high. The detailed results can be found in Fig. 6 (fault diagnosis result), Table VIII (model complexity), and Table IX (inference time) of section IV-D.

\section{Experiments Result}

\subsection{Data Description}
The Padderborn University bearing database is an open database released by Padderborn University, which provides vibration signal data according to the operation of an electric motor collected using the device in Fig. 6 \cite{lessmeier2016condition}. In this study, the vibration signal collected through this test device was used to verify the performance of a fault diagnosis method.
\begin{figure}[ht]
    \centering
    \includegraphics[width=\columnwidth]{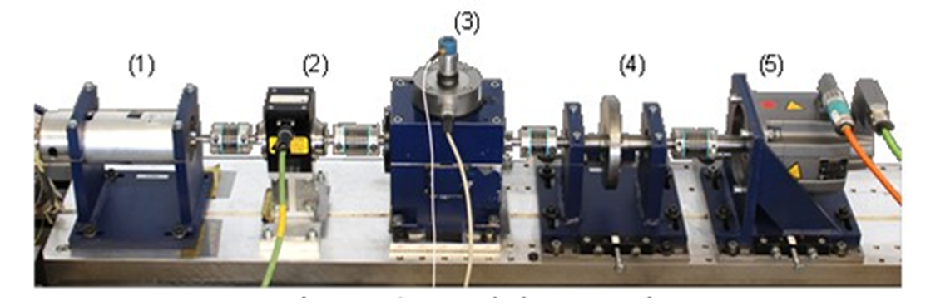}
    \caption{Paderborn University bearing data test rig.}
    \label{fig:test_rig}
\end{figure}

In the Paderborn University vibration dataset, there are normal data and two types of abnormal data: i) outer rate fault, ii) inner race fault. In addition, as shown in Table \ref{tab:working condition}, four operating environments were considered.

\begin{table}[!h]
    \caption{Divide dataset for each working condition}
    \label{tab:working condition}
    \centering
    \small
    \resizebox{0.8\columnwidth}{!}{
    \begin{tabular}{c|ccccc}
         \hline\hline
         Dataset & RPM & Load Torque & Radial force \\
         \hline
         A       & 1500 & 0.7 & 1000 \\ 
         B       & 900  & 0.7 & 1000  \\
         C       & 1500 & 0.7 & 400   \\
         D       & 1500 & 0.1 & 400   \\
         \hline\hline
    \end{tabular}
    }
\end{table}

In this paper, in addition to these four operating environments, two types of abnormal state data were considered according to normal data and inner race and outer race. In the case of abnormal data, the actual collected data were used for model learning and performance verification.
Five different bearings were considered for each of the three classes classified according to the normal and abnormal cause location. Table \ref{tab:bearing set} summarizes the datasets used for verification in this paper.

\begin{table}[h]
    \caption{Bearing dataset categorization for bearing condition state and damage type used in the experiment.}
    \label{tab:bearing set}
    \centering
    \small
    \resizebox{0.8\columnwidth}{!}{
    \begin{tabular}{c|c}
         \hline\hline
         Bearing Dataset & Bearing Condition State \\
         \hline
        \makecell[c]{K001, K002, K003, \\K004, K005} &  Normal \\
        \hline
        \makecell[c]{KA04, KA15, KA16, \\KA22, KA30} &  Outer Race Fault \\
        \hline
        \makecell[c]{KI04, KI14, KI16, \\KI18, KI21} &  Inner Race Fault \\
        \hline
        KB23, KB24, KB27 & \makecell[c]{Inner and Outer \\ Race Fault}  \\
        \hline\hline
    \end{tabular}
    }
\end{table}

\subsection{Data Preprocessing}
The Paderborn University dataset provides current data about 79 to 80 files for each bearing, and each file contains $256,000$ sampled data values. The Paderborn University bearing set has four operating environments for each bearing, and 79-80 sampled data exist for each bearing code. Each sampled data is sampled at 64 kHz per second. In this paper, segmentation was performed as follows to assume that the machine diagnoses the defect every time it rotates.

\begin{align}
    \label{eq:segment}
    L=Q \cdot f_s \cdot RPS^{-1},
\end{align}

Here, RPS is rotation per second (RPS) = $RPM/60$, and $f_s$ is the sampling rate per second. There are two RPM in the Paderborn University bearing set, and in this paper, segmentation was performed based on 900 RPM to consider the speed of the slowest rotating machine. The performance verification for each signal-to-noise ratio (SNR) was performed by applying noise to the segmented pure vibration signal and pure vibration signal. It was verified using two noises (Gaussian, Laplace noise) to verify whether it is a suitable model in various work environments. Data preprocessing was performed by synthesizing the above two noises into the pure signal. SNR is calculated as follows:

\begin{align}
    SNR = 10 \times log_{10}(\frac{P_{signal}}{P_{noise}})
\end{align}

Here, $P_{noise}$ means the strength of noise, and $P_{signal}$ means the strength of signal. The range of SNR used in this paper is from -8 dB to 6 dB, and a total of eight SNR were used for verification at 2 dB intervals.

Train: validation: Test data segmentation ratio was set to 8:1:1 for learning, verifying, and testing the segmentation and data preprocessed dataset. The model loss was cross entry loss, the optimizer was AdamW, and the weight decay ratio was 0.00001.

In this paper, the model was designed and trained based on the Pytorch framework, and the workstation consists of CPU Intel(R) i9-10900X, 256GB of RAM, and GPU NVIDIA GeForce RTX 3090 Ti$\times$2. The edge device used Jetson Nano.

\subsection{Proposed Network Fault Diagnosis Result}
In this section, fault diagnosis accuracy verification was conducted to confirm that the proposed LSR-Net is a model capable of robust fault diagnosis according to operating environment, type of noise, and SNR. In addition, in order to determine whether the use of DFEM had anti-noising and feature enhancement effects, a comparison of fault diagnosis accuracy was conducted with and without the use of DFEM.

\subsubsection{Performance Comparison by Channel split Ratio}
Table \ref{tab:Channel_split} is the result of comparing the fault diagnosis accuracy when the channel split ratio of the CES block is changed, the calculation complexity of the model, and the inference time. The channel split ratio is expressed in the order of $\mathbf{X}^{left}:\mathbf{X}^{right}$. The experiment was conducted. As can be seen in Table IV, the fault diagnosis accuracy was the highest when the Conv operation was performed more at 3:7 in the CES block. On the other hand, when the ratio of the channel used for the Conv operation was reduced to 7:3, it was effective in terms of the computational complexity of the model. Accordingly, in this paper, a 5:5 ratio of channel split ratio was adopted to consider the trade-off between the fault diagnosis accuracy and the computational complexity of the model.

\begin{table}[!h]
    \caption{Comparison of various metric according to channel split ratio}
    \label{tab:Channel_split}
    \centering
    {\large
    \resizebox{\columnwidth}{!}{
        \begin{tabular}{c | c c c}
             \hline\hline
             \diagbox[width=8em]{Metric}{Ratio} & 7:3 & 5:5 & 3:7 \\
             \hline
             Accuracy & $ 99.934 \pm 0.06$ & $99.970 \pm 0.03$ & $\mathbf{99.973 \pm 0.04}$ \\ [0.5ex]
             FLOPs ($10^6$) & $\mathbf{0.712}$ & $0.717$ & $0.722$ \\ [0.5ex]
             Parameters ($10^3$) & $0.671$ & $\mathbf{0.613}$ & $0.732$ \\ [0.5ex]
             Inference time (ms) & $\mathbf{3.280 \pm 0.42}$ & $3.372 \pm 0.55$ & $3.403 \pm 0.73$ \\[0.5ex]
            \hline\hline
        \end{tabular}
    }}
\end{table}

\subsubsection{Performance Comparsion by DFEM Methods}

Table \ref{tab:DFEM} shows the results of comparing the defect diagnosis accuracy depending on whether DFEM is used or not. For models that did not use DFEM, the raw signal was converted to a 2D matrix and used as input data for the model. For defect diagnosis accuracy, the average accuracy for each noise type and SNR was compared. As can be seen in Table VI, when DFEM was used, the defect diagnosis accuracy was confirmed to be 4.566\% higher on average than when DFEM was not used. Additionally, it was confirmed that when FEM was used, the defect diagnosis accuracy was 0.749\% higher on average than when FEM was not used. Therefore, it was confirmed that the use of DFEM can improve the denoising and feature enhancement capabilities of the model.

\begin{table}[!hbt]
    \caption{Comparison of bearing fault diagnosis accuracy according to DFEM methods}
    \label{tab:DFEM}
    \centering
    {\large
    \resizebox{\columnwidth}{!}{
        \begin{tabular}{c | c c c}
             \hline\hline
             \diagbox[width=8em]{Method}{Noise} & Gaussian & Laplace & Average \\
             \hline
             Proposed Methods & $\mathbf{98.135 \pm 2.24}$ & $\mathbf{98.052 \pm 1.89}$ & $\mathbf{98.094 \pm 1.94}$ \\ [0.5ex]
             Without DFEM     & $93.576 \pm 5.25$ & $93.481 \pm 5.38$ & $93.528 \pm 5.14$ \\ [0.5ex]
             Without FEM       & $97.574 \pm 2.63$ & $97.116 \pm 3.12$ & $97.345 \pm 2.66$ \\ [0.5ex]
            \hline\hline
        \end{tabular}
    }}
\end{table}

\begin{figure}[!hbt]
    \centering
    \includegraphics[width=\columnwidth]{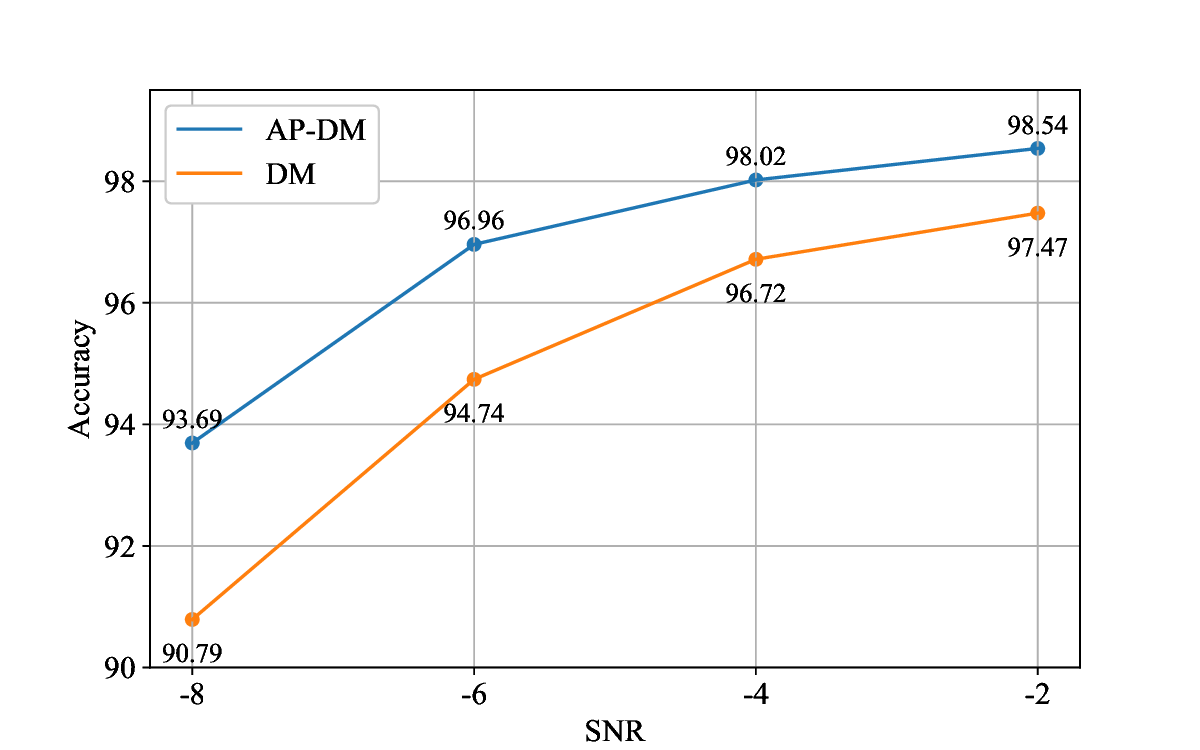}
    \caption{Comparison of negative SNR fault diagnosis accuracy by adaptive pruning in DM.}
    \label{fig:AP-DM_and_DM}
\end{figure}

Fig. \ref{fig:AP-DM_and_DM} is the result of comparing the average defect diagnosis accuracy of Gaussian and Laplace noise in a situation with negative SNR (-8 to -2) when using AP-DM rather than DM. Fig. As can be seen in Fig. \ref{fig:AP-DM_and_DM}, when learning is conducted using AP, the stronger the noise power, the more effective it is. Through this, since learning is performed on noise that has a random effect when it is greatly affected by noise, the application of pruning can improve generalization performance from noise data by randomly removing weights that may cause overfitting. Confirmed. If noise and signal power are measured in an actual work environment and are greatly affected by noise, it will be possible to improve anti-noise ability by using AP-DM.

\vspace{-10pt}

\begin{table*}[!ht]
    \caption{Comparison of bearing fault diagnosis accuracy for proposed model by SNR in Gaussian noise}
    \label{tab:Gaussian_noise}
    \centering
    {\small
    \resizebox{1.5\columnwidth}{!}{
        \begin{tabular}{c c c c c c}
             \hline\hline
             \diagbox[width=8em]{SNR}{Dataset} & A & B & C & D & Average \\
             \hline
             pure\,signal & $100 \pm 0$ & $100 \pm 0$ & $100 \pm 0$ & $100 \pm 0$ & $100 \pm 0$\\ [0.5ex]
             -8           & $97.166 \pm 0.39$ & $95.653 \pm 0.88$ & $96.096 \pm 1.07$ & $99.096 \pm 0.45$ & $96.503 \pm 0.96$\\[0.5ex]
             -6           & $97.971 \pm 0.62$ & $97.027 \pm 0.91$ & $98.123 \pm 0.54$ & $97.556 \pm 0.15$ & $97.669 \pm 0.78$\\[0.5ex]
             -4           & $97.896 \pm 0.72$ & $97.061 \pm 0.89$ & $98.134 \pm 0.36$ & $97.942 \pm 0.13$ & $97.758 \pm 0.74$ \\[0.5ex]
             -2           & $98.765 \pm 0.19$ & $97.613 \pm 1.09$ & $98.413 \pm 0.26$ & $98.628 \pm 0.12$ & $98.355 \pm 0.70$\\[0.5ex]
              0           & $99.592 \pm 0.06$ & $98.167 \pm 0.32$ & $99.442 \pm 0.15$ & $99.582 \pm 0.06$ & $99.446 \pm 0.24$\\[0.5ex]
             2            & $99.958 \pm 0.05$ & $99.493 \pm 0.09$ & $99.892 \pm 0.06$ & $99.980 \pm 0.03$ & $99.831 \pm 0.21$\\[0.5ex]
             4            & $99.938 \pm 0.03$ & $99.722 \pm 0.06$ & $99.946 \pm 0.08$ & $99.980 \pm 0.03$ & $99.897 \pm 0.12$\\[0.5ex]
             6            & $100\pm0$ & $99.921\pm0.05$ & $99.99\pm0.02$ & $99.968 \pm 0.05$ & $99.970 \pm 0.05$\\
             
            \hline\hline
        \end{tabular}
    }}
\end{table*}
\begin{table*}[!ht]
    \caption{Comparison of bearing fault diagnosis accuracy for proposed model by SNR in Laplace noise}
    \label{tab:Laplace_noise}
    \centering
    {\small
    \resizebox{1.5\columnwidth}{!}{
        \begin{tabular}{c c c c c c}
             \hline\hline
             \diagbox[width=8em]{SNR}{Dataset} & A & B & C & D & Average \\
             \hline
             pure\,signal & $100 \pm 0$ & $100 \pm 0$ & $100 \pm 0$ & $100 \pm 0$ & $100 \pm 0$\\ [0.5ex]
             -8           & $96.758\pm0.78$ & $96.610\pm1.40$ & $96.761\pm0.39$ & $97.642\pm0.48$ & $97.096\pm0.85$\\ [0.5ex]
             -6           & $97.982\pm0.70$ & $96.565\pm1.09$ & $96.965\pm0.66$ & $98.124\pm0.53$ & $97.471\pm1.01$\\ [0.5ex]
             -4           & $98.154\pm0.42$ & $96.475\pm0.73$ & $97.845\pm0.55$ & $98.703\pm0.15$ & $97.856\pm0.97$\\ [0.5ex]
             -2           & $98.830\pm0.29$ & $97.916\pm0.82$ & $98.048\pm0.54$ & $99.928\pm0.24$ & $98.267\pm0.66$\\ [0.5ex]
             0            & $99.722\pm0.04$ & $99.232\pm0.14$ & $99.475\pm0.21$ & $99.732\pm0.04$ & $99.540\pm0.24$\\ [0.5ex]
             2            & $99.925\pm0.06$ & $99.561\pm0.20$ & $99.775\pm0.07$ & $99.883\pm0.10$ & $99.786\pm0.18$\\ [0.5ex]
             4            & $99.959\pm0.04$ & $99.639\pm0.11$ & $99.903\pm0.04$ & $99.970\pm0.03$ & $99.868\pm0.15$\\ [0.5ex]
             6            & $100\pm0$       & $99.764\pm0.10$ & $99.893\pm0.04$ & $99.937\pm0.07$ & $99.898\pm0.11$\\ [0.5ex]
             
            \hline\hline
        \end{tabular}
    }}
\end{table*}

\subsubsection{Gaussian and Laplace Noise Fault Diagnosis Result}

Table VI shows the results of comparing defect diagnosis accuracy using Gaussian noise and Table VII uses Laplace noise. In the case of pure signal, it was confirmed that accurate defect diagnosis of 99.977\% of the dataset was possible. In addition, it was confirmed that even if affected by strong noise of -8dB, robust defect diagnosis is possible with an average of 93.273\% in the case of Gaussian noise and 94.112% in the case of Laplace noise, depending on the type of noise.

Table \ref{tab:Gaussian_noise} is the result of comparing defect diagnosis accuracy using Gaussian noise, and Table \ref{tab:Laplace_noise} is Laplace noise. In the case of pure signals, perfect fault diagnosis accuracy of 100\% on dataset average was confirmed. In addition, even when affected by strong noise of -8dB, high fault diagnosis accuracy was confirmed, depending on the type of noise, with an average of 96.5\% for Gaussian noise and 97.1\% for Laplace noise.

\subsection{Comparsion with Benchmark Model}
In this paper, CA-CNN \cite{cui_22}, LEFE-Net \cite{fang_21b}, SIRCNN \cite{yao_20}, MCDS-CNN \cite{ling_22}, and SE-IRCNN \cite{he_21}, which conducted bearing defect diagnosis research using existing compact model design and SNR, were selected as benchmark models and compared defect diagnosis accuracy, model complexity, inference time, and power consumption by SNR. In addition, the number of LSR-Net and filters proposed in this paper during the Conv operation is the same, but channel split and channel shuffle are not used, and DSC-Net designed using DSConv and PConv instead of the general Conv and GPConv was also benchmarked and compared. The defect diagnosis accuracy was compared by using the average defect diagnosis accuracy for each SNR in all operating environments (dataset A-D). The defect diagnosis accuracy was compared by using the average defect diagnosis accuracy for each SNR in all operating environments (dataset A-D).

\subsubsection{Gaussian and Laplace Noise fault diagnosis result}

\begin{figure*}[!hbt]
    \centering
    \includegraphics[width=1.8\columnwidth]{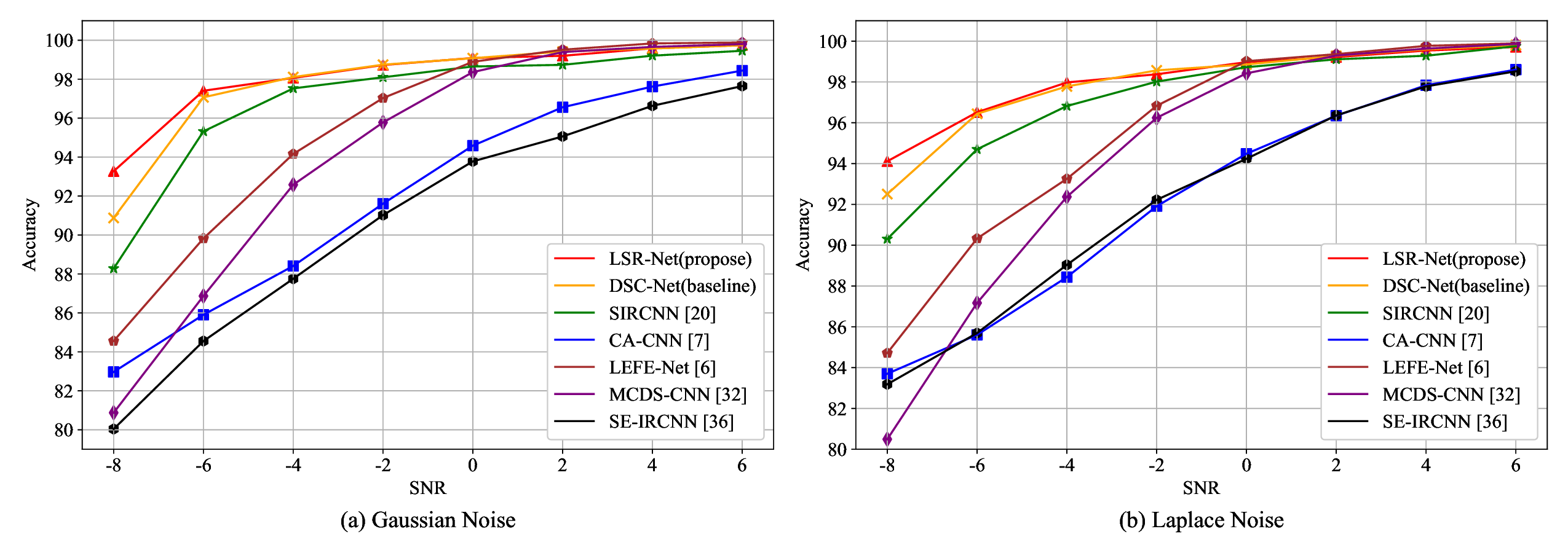}
    \caption{Comparison of the accuracy of bearing fault diagnosis by SNR ( (a) Gaussian and (b) Laplace noise).}
    \label{fig:benchmark_acc}
\end{figure*}

Fig. \ref{fig:benchmark_acc} is the result of comparing the defect diagnosis accuracy by SNR of Gaussian and Laplace noises of each model. As can be seen from Fig. \ref{fig:benchmark_acc}, in the case of LSR-Net and DSC-Net, through the denoising and feature enhancement process using DFEM, it was confirmed that high defect diagnosis accuracy can be maintained in all SNRs even if they are heavily affected by noise because the effect on noise was small. On the other hand, in the case of existing benchmark models, it can be seen that the more affected by noise, the more the defect diagnosis accuracy decreases.

Table \ref{tab:total_acc} is the average defect diagnosis result in pure signal and Gaussian, Laplace noise of the proposed model and benchmark models. As a result of comparing the defect diagnosis accuracy of DSC-Net using DSConv also proposed by Pure signal with the proposed LSR-Net, it was confirmed that LSR-Net can diagnose defects 0.18\% more accurately in Gaussian noise and 0.12\% in Laplace noise than in DSC-Net.

\begin{table}[!h]
    \caption{Comparison of the accuracy of bearing fault diagnosis by SNR}
    \label{tab:total_acc}
    \centering
    {\large
    \resizebox{0.9\columnwidth}{!}{
        \begin{tabular}{c | c c c}
             \hline\hline
             \diagbox[width=7em]{Model}{Noise} & Pure Signal & Gaussian & Laplace \\
             \hline
             \makecell[c]{LSR-Net \\ (propose)} & $99.977 \pm 0.03$ & $\mathbf{98.135 \pm 2.24}$ & $\mathbf{98.052 \pm 1.89}$ \\ [0.5ex]
            \makecell[c]{DSC-Net  \\ (baseline)} & $99.959 \pm 0.03$ & $97.824 \pm 2.94$ & $97.854 \pm 2.42$ \\ [0.5ex]
            \makecell[c]{SIRCNN \\ \cite{yao_20}} & $99.977 \pm 0.03$ & $96.911 \pm 3.72$ & $97.068 \pm 3.60$ \\ [0.5ex]
            \makecell[c]{CA-CNN  \\ \cite{cui_22}}  & $99.641 \pm 0.25$ & $92.012 \pm 5.76$ & $92.115 \pm 5.67$ \\ [0.5ex]
            \makecell[c]{LEFE-Net \\ \cite{fang_21}} & $99.939 \pm 0.03$ & $95.462 \pm 5.63$ & $95.396 \pm 5.54$ \\ [0.5ex]
            \makecell[c]{MCDS-CNN \\ \cite{ling_22}} & $\mathbf{99.986 \pm 0.01}$ & $94.164 \pm 6.99$ & $94.185 \pm 7.07$ \\ [0.5ex]
             \makecell[c]{SE-IRCNN \\ \cite{he_21}} & $99.743 \pm 0.12$ & $90.810 \pm 6.23$ & $92.129 \pm 5.68$ \\ [0.5ex]
            \hline\hline
        \end{tabular}
    }}
\end{table}

In this experiment, the fault diagnosis accuracy of the model was verified in two environments: Gaussian and Laplace noise. In the case of the proposed LSR-Net, the average difference in defect diagnosis accuracy between Gaussian and Laplace noise was 0.04\%, confirming that it is an effective defect diagnosis model regardless of the type of noise. On the other hand, SIRCNN has a large difference in defect diagnosis accuracy of 6.26\% on average depending on the type of noise.

\subsubsection{Model Complexity}
In this section, the computational complexity of the model was compared using the proposed LSR-Net and the benchmark models' FLOPs, parameters, and model size because the purpose of this section is to design a model that can smoothly diagnose a fault in an edge device where bearing defect diagnosis is mainly performed. The results can be found in Table \ref{tab:model_complexity}.

\begin{table}[!h]
    \caption{Comparison with each model FLOPs, Parameters, and Model Size}
    \label{tab:model_complexity}
    \centering
    {\small
    \resizebox{0.8\columnwidth}{!}{
        \begin{tabular}{c | c c c}
             \hline\hline
             \diagbox[width=8em]{Model}{Metric} & \makecell[c]{FLOPs \\ $(10^6)$} & \makecell[c]{Parameters \\ $(10^6)$} & \makecell[c]{Model Size \\ (KB)} \\
             \hline
             \makecell[c]{LSR-Net \\ (propose)}       & $\mathbf{0.716}$ & $\mathbf{0.613}$ & $\mathbf{53.50} $ \\ [0.5ex]
             \makecell[c]{DSC-Net  \\ (baseline)}     & $0.756$ & $1.594$ & $89.90 $ \\ [0.5ex]
             \makecell[c]{SIRCNN \\ \cite{yao_20}}    & $281.4$ & $27.25$ & $1090.0$ \\ [0.5ex]
             \makecell[c]{CA-CNN  \\ \cite{cui_22}}   & $2.919$ & $41.17$ & $1760.0$ \\ [0.5ex]
             \makecell[c]{LEFE-Net \\ \cite{fang_21}} & $6.550$ & $7.066$ & $472.0 $ \\ [0.5ex]
             \makecell[c]{MCDS-CNN \\ \cite{ling_22}} & $1.129$ & $6.950$ & $286.0 $\\ [0.5ex]
             \makecell[c]{SE-IRCNN \\ \cite{he_21}}   & $1.904$ & $3.248$ & $250.0 $\\ [0.5ex]
             \makecell[c]{LiteFormer \\ \cite{sun_23}} & $123.1$ & $241.0$ & $976.0$ \\ [0.5ex]
            \hline\hline
        \end{tabular}
    }}
\end{table}

As can be seen from Table \ref{tab:model_complexity}, it was confirmed that the FLOPs, Parameters, and model size of LSR-Net were very low compared to the benchmark model. Among them, when comparing the computational complexity of SIRCNN and the model, it was confirmed that there was a big difference of about 392 times for FLOPs, about 35 times for parameters, and about 18 times for model size. Since LSR-Net was designed based on GConv and GPConv, it was possible to confirm the computational complexity of a smaller model than the DSC-Net that designed the model using DSConv.

\subsubsection{Inference Time}

In this section, the proposed model and benchmark models were compared by measuring the inference time using the target platform edge device. The inference time measurement method measured the average inference time repeated 128 times to check whether there is a model with a large load when inference was repeated. In addition, in order to measure the inference time in a more restrictive environment, the experiment was conducted by adjusting the number of CPU clocks used in the edge device. The results can be found in Table \ref{tab:inference_time}. As can be seen from Table \ref{tab:inference_time}, LSR-Net was able to confirm the fastest inference time and the lowest delay time in the edge environment. In addition, it was confirmed that fast inference is possible even when the number of CPU clocks is reduced due to the limited power of the edge device.

\begin{table*}[ht]
    \caption{Comparison with each model inference time in desktop and edge device using CPU}
    \label{tab:inference_time}
    \centering
    {\small
    \resizebox{1.8\columnwidth}{!}{
        \begin{tabular}{l c c c c c c c}
             \hline\hline 
             \diagbox[width=8em]{Time (ms)}{Model} & \makecell[c]{LSR-Net \\ (propose)} 
             & \makecell[c]{DSC-Net \\(baseline)}
             & \makecell[c]{SIRCNN \\ \cite{yao_20}}
             & \makecell[c]{CA-CNN \\ \cite{cui_22}} & \makecell[c]{LEFE-Net \\ \cite{fang_21}}
             & \makecell[c]{MCDS-CNN \\ \cite{ling_22}} & \makecell[c]{SE-IRCNN \\ \cite{he_21}} \\
             \hline
             \makecell[c]{Edge device \\ (4 Clocks)} & $\mathbf{10.29 \pm 4.23}$ & $16.56 \pm 10.1$ & $439.37 \pm 65.8$ & $183.72 \pm 41.0$ &	$43.25 \pm 14.53$ & $29.29 \pm 11.31$ & $41.39 \pm 11.54$ \\ [1.5ex]
            \makecell[c]{Edge device \\ (2 Clocks)} & $\mathbf{18.16 \pm 6.77}$ & $29.62 \pm 10.2$ &	$1032.2 \pm 526$ & $246.06 \pm 80.9$ &	$53.432 \pm 15.9$ &	$40.09 \pm 15.08$ &	$61.25 \pm 20.05$ \\
            \hline\hline
        \end{tabular}
    }}
\end{table*}
Moreover, when DSC-Net with a high MAC/FLOPs ratio was verified in the edge device using many groups without applying the channel split to the LSR-Net, the inference time was slower than that of the LSR-Net. Therefore, it was confirmed that the number of groups and channels used in the Conv operation are important when designing a lightweight model. 

Accordingly, the LSR-Net may be capable of real-time defect diagnosis using faster inference time and lower delay time than the benchmark models in the edge device. Moreover, it may be confirmed that the inference time of the LEFE-Net is shorter in the case of the CA-CNN even though it has lower FLOPs than that of the LEFE-Net. Therefore, it may not be appropriate to design a model using only indirect indicators.

\subsubsection{Power Consumption}

\begin{figure}[h]
    \centering
    \includegraphics[width=\columnwidth]{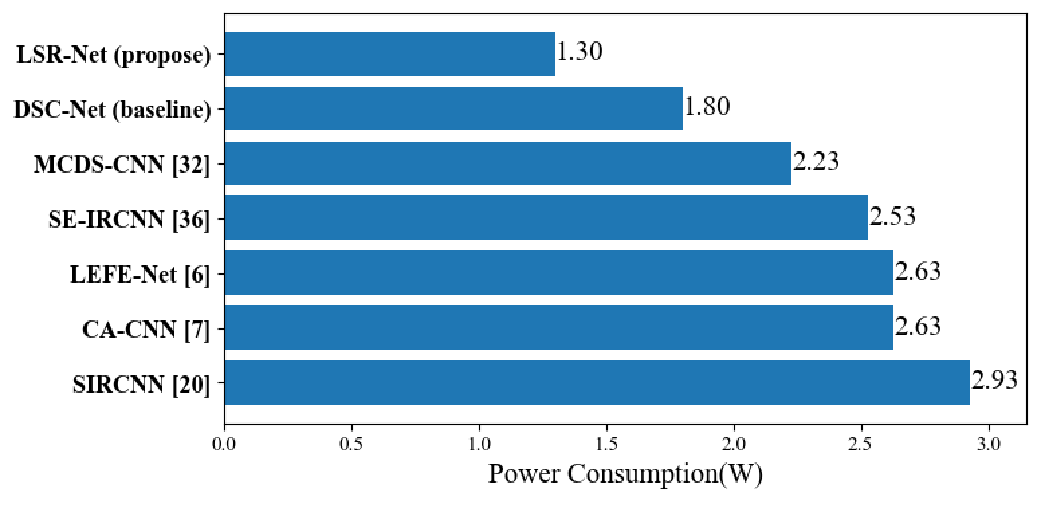}
    \caption{Comparison with each model power consumption}
    \label{fig:power_consumtion}
\end{figure}

High power consumption during model operation can cause edge device overload, resulting in poor edge device efficiency, resulting in latency during defect diagnosis \cite{garland_18,tang_20}. Therefore, as shown in Fig. \ref{fig:power_consumtion}, the power consumption (W) consumed when the model operates at the edge device was compared. As can be seen from Fig. \ref{fig:power_consumtion}, the model was designed without using exponential computation, and it was confirmed that the LSR-Net with very small model complexity of the model had the lowest power consumption. As a result, a lower standard deviation was confirmed compared to other benchmark models when inferring from the edge device.

\vspace{2\baselineskip}

\section{Conclusion}
In this paper, we proposed an LSR-Net that is accurate even in noise-affected environments and is capable of diagnosing defects in real time. It was confirmed that robust defect diagnosis is possible through denoising using 1D Conv and adaptive pruning and feature enhancement using activation ensemble even if the collected signal is affected by noise by using DFEM in the LSR-Net. In addition, it was possible to design the model while maintaining the low computational complexity of the model by characterizing the feature-map that passed through the DFEM using the CES block designed in consideration of the model's computational complexity and MAC/FLOPs ratio compared to the existing DSConv. As a result, even when affected by noise compared to the existing benchmark model, high defect diagnosis accuracy of 98.68-98.72\% was maintained on average, and a low delay of about 4 ms and fast inference of 10 ms were possible in the edge device. Through this, effective bearing defect diagnosis will be possible in actual workplaces that are affected by noise and require real-time performance.

% \section*{Acknowledgments}
% This should be a simple paragraph before the References to thank those individuals and institutions who have supported your work on this article.


\bibliographystyle{IEEEtran}
\bibliography{reference}

\vfill

\end{document}